\documentstyle{article}
\begin{document}

\begin{center}
DIPOLE RESONANCES AND THE NUCLEAR SCHIFF MOMENT

\vspace{0.5cm}
Naftali Auerbach$^{1,2}$ and Vladimir Zelevinsky$^{2,3}$ \\
\vspace{0.5cm}
$^{1}${\sl School of Physics and Astronomy, Tel Aviv University, Tel Aviv, 69978, Israel \\
$^{2}$Department of Physics and Astronomy, Michigan State University, \\
East Lansing, Michigan 48824-2320, USA\\
$^{3}$National Superconducting Cyclotron Laboratory, Michigan State University,
East Lansing, Michigan 48824-1321, USA}
\end{center}


\begin{abstract}
The nuclear Schiff moment creates a mechanism of transfer of the violation of parity and time-reversal
invariance by weak interaction in nuclei into the atomic electric dipole moment. We point out an
additional contribution to the Schiff moment generated by the mixing of single-particle states
through the low-lying nuclear dipole resonances. An estimate shows that this contribution is by order
of magnitude comparable to single-particle contributions and can be enhanced if the low-lying resonance
has collective nature.
\end{abstract}

\section{Introduction}

There is a close connection between collective nuclear motion and symmetries in the nuclear Hamiltonian. In some instances collective excitations are the result of the existence of certain symmetries in the system. One of the best examples is the isobaric analog resonance that results from the charge symmetry of the nuclear force. In other instances however collective excitations, such as giant resonances, serve as intermediate states in the process of breaking symmetries. Again, when we consider the isospin symmetry, the giant isovector monopole state is the important intermediate state that facilitates the breaking of this symmetry by the Coulomb interaction. Other well known examples are $J^{\pi}=0^{-}$ component of the spin-dipole giant resonance that plays a role in parity mixing; the quadrupole, $J^{\pi}=2^{+}$, together with the octupole, $J^{\pi}=3^{-}$, excitations contribute to the enhancement of the violation of time reversal symmetry when a ${\cal T}$-odd (time reversal odd) and ${\cal P}$-odd (parity odd) interaction is present. Some nuclei experience a phase transition when the ground states become quadrupole
and/or octupole deformed as a result of the lowering of the quadrupole and octupole modes. In the case of the combined quadrupole+octupole deformation, or corresponding soft vibrations in spherical nuclei, the ${\cal T}$- and ${\cal P}$-odd Schiff moment (see next section for the definition) becomes strongly enhanced. The Schiff moment induces
a ${\cal T}$-odd field that in turn produces an atomic electric dipole moment.

In recent years attention was given to the study of new nuclear resonances at relatively low energy and to some specific properties of resonances known before. For example, concentration of the low-lying isovector dipole ($J^{\pi}=1^{-},\; T=1$) strength was observed in many nuclei, and termed the Pygmy Dipole resonance (PDR), see
\cite{tonchev10,endres12} and references therein. A resonance that carries the quantum numbers $J^{\pi}T=1^{-}0$, the so-called isoscalar dipole resonance (ISDR), sometimes called also ``compression" or ``squeezing" mode, was studied extensively experimentally \cite{morsch80,morsch83,davis97,garg04} with the low component observed in many nuclei, see for example \cite{clark01,uchida04,nayak06}. General
theoretical estimates based on the sum rules were performed long ago \cite{giai81,harakeh81}.  As in the case of the isovector dipole, one finds that the strength is split into two energy regions, a low-energy concentration of strength and a ``main peak" at higher energy. This comes naturally in the quasiparticle random phase approximation
and similar approaches with different assumptions concerning mean field and residual interactions
\cite{colo00,piekarewicz00,shlomo02,paar07,vretenar12}.

The characteristic mechanism of appearance of low-energy strength concentration can be illustrated by the random phase approximation: the coherent interaction, usually of the multipole-multipole type, separates (up in energy for the isovector channels) a ``giant" collective superposition of particle-hole excitations with appropriate quantum numbers; however, the remaining strength is still located around the shell-model excitation energy and can be also collectivized up to some extent, depending on the specific nuclear properties. In particular, the collectivity can be enhanced in loosely bound nuclei with a noticeable neutron excess filling the outer orbitals and vibrating with respect to the core.
If the low-lying excitations are partly above the neutron separation threshold, the collectivization through continuum also can increase the resonance strength \cite{Szagreb}. The isoscalar dipole resonance is of particular interest to us here because the corresponding exciting operator is identical to the Schiff operator involved in the time reversal studies of the atomic electric dipole moment. Because of this close connection between the ISDR and the Schiff operator it is important to use the measured (or calculated) ISDR strength to evaluate Schiff moments in nuclei.

In what follows we consider the emergence of the non-zero Schiff moment in a spherical odd-$A$ nucleus as a result of the weak interaction that can mix states of opposite parity through admixture of the ISDR excitation
of a neighboring even core. In contrast to the known quadrupole-octupole mechanisms here we do not expect large enhancement due to the parity doublets and small energy denominators. However, the presence of the collective strength
makes such contributions at least comparable to the effects typically taken into account in the available calculations
on a pure single-particle level. This means that such coupling has to be taken into account in microscopic consideration.

\section{Mixing of states by weak interaction}

We consider two neighboring nuclei, an even-even one with the ground state $|0^{+}\rangle$
of angular momentum $J=0$ and positive parity and its odd neighbor with an unpaired
nucleon in the spherical state $|jm;+\rangle$, let say also of positive parity
(just for definiteness, this is of no importance),
\begin{equation}
|jm;+\rangle=a^{\dagger}_{jm}|0^{+}\rangle.                  \label{1}
\end{equation}
To illustrate our idea we use here the simplest shell-mode wave functions.
Both nuclei are supposed to have relatively low-lying dipole excitations with typical excitation
energy $\Delta E$ up to 10 MeV. Because of negative parity, these excited states are concentrated
in the region of the next oscillator shell and may form a collective quasicontinuum, the so-called
{\sl pygmy resonance}. Although the main part of the isovector dipole strength is shifted up
to the region of the giant dipole resonance, here we still have a significant leftover, not
necessarily in the form of a single collectivized RPA-type excitation. It is expected that this part of
dipole strength can be more noticeable for nuclei away from the valley of stability; it can be
of isovector or isoscalar character.

Due to the angular momentum coupling, the dipole states $|1^{-}\mu\rangle$ of the even nucleus
with angular momentum $J=1$, its projection $J_{z}=\mu$ and  negative parity have their
cousins in the odd nucleus with the same angular momentum quantum numbers as the odd-$A$ ground state,
\begin{equation}
|jm;-\rangle=\sum_{\mu m'}C^{jm}_{1\mu\;jm'}a^{\dagger}_{jm'}|1^{-}\mu\rangle;     \label{2}
\end{equation}
here and below we use the Clebsch-Gordan coefficients of vector coupling.
Now we have the two states, eqs. (\ref{1}) and (\ref{2}), in the same
(odd) nucleus with the same angular momentum and opposite parity. The parity-violating
weak interaction $W$ will mix these states creating their linear combinations,
\begin{equation}
|jm\rangle_{0}=|jm;+\rangle -\alpha|jm;-\rangle,                            \label{3}
\end{equation}
for the new ground state and
\begin{equation}
|jm\rangle_{1}=|jm;-\rangle+\alpha^{\ast}|jm;+\rangle                      \label{4}
\end{equation}
for the new excited state. The admixture amplitude is determined by the corresponding matrix
element $W_{10}$ of the weak interaction,
\begin{equation}
\alpha=\,\frac{W_{10}}{\Delta E}.                                      \label{5}
\end{equation}
These equations are trivially generalized for the case of several admixed excited states.

Since the mean value of the nuclear dipole moment ${\bf D}$ is screened in the atom, the atomic
electric dipole moment is generated by the nuclear {\sl Schiff moment} [see the recent reexamination
of the Schiff theorem in \cite{senkov08,AZschiff08}],
\begin{equation}
{\bf S}=x\sum_{a}e_{a}\left(r_{a}^{2}-\,\frac{5}{3}\langle r^{2}\rangle_{{\rm ch}}\right){\bf r}_{a}, \label{6}
\end{equation}
where $x=1/10$ in the standard definition, and $\langle r^{2}\rangle_{{\rm ch}}$ is the mean square charge radius,
so that, with the radius $R_{p}$ of the equivalent homogeneous sphere, $(5/3)\langle r^{2}\rangle_{{\rm ch}} =R_{p}^{2}$. This operator with $x=1$ and for $e_{a}=1$ is identical to the ISDR operator that appears in
the long wavelength expansion of spherical wave with $\ell=1$ (the main dipole component that leads to the spurious
excitation of the center of mass is excluded).

The operator (\ref{6}) can have a non-vanishing expectation value in the mixed ground state (\ref{3}),
\begin{equation}
_{0}\langle jm'|S_{\kappa}|jm\rangle_{0}=-\,\frac{2\,{\rm Re}\,\alpha}{\sqrt{2j+1}}\,
C^{jm'}_{jm\;1\kappa}\,(j;+||S||j;-),                                  \label{7}
\end{equation}
where $S_{\kappa}$ are the spherical components of the vector ${\bf S}$ and we use the definition of reduced
(double-barred) matrix elements according to Edmonds \cite{edmonds57}.

The matrix elements of the Schiff moment between the states of opposite parity in the odd nucleus
are given by
\begin{equation}
\langle jm';+|S_{\kappa}|jm;-\rangle=\langle 0^{+}|a_{jm'}S_{\kappa}\sum_{m''\mu}
C^{jm}_{1\mu\;jm''}\,a^{\dagger}_{jm''}|1^{-}\mu\rangle             \label{8}
\end{equation}
Here the collective enhancement can emerge from the excitation of the even core through the low-lying
dipole modes. This contribution corresponds to $m''=m'$ and therefore $\mu=-\kappa$;
with the same definition of reduced matrix elements,
\begin{equation}
\langle jm';+|S_{\kappa}|jm;-\rangle\approx \,\frac{1}{\sqrt{3}}\,(-)^{1-\kappa}C^{jm}_{1-\kappa\;jm'}
(0^{+}||S||1^{-}).                                                   \label{9}
\end{equation}
This amplitude, possibly summed over several dipole excitations, can be of collective nature.

\section{Effective weak interaction}

The effective ${\cal P}$- and ${\cal T}$-violating interaction can be introduced in various forms.
In our space of states, the effective form should look like
\begin{equation}
W=\xi({\bf D}\cdot\vec{\sigma}),                                \label{10}
\end{equation}
where $\xi$ is the effective coupling constant, ${\bf D}$ is a coordinate-dependent ${\cal P}$-odd
${\cal T}$-even vector operator,
such as a dipole moment or Schiff moment, creating collective excitations in the even core, while
the spin operator $\vec{\sigma}$ acts on the odd particle. The mixing (\ref{5}) is determined by the matrix element
\begin{equation}
W_{10}\;\Rightarrow\;\langle jm';-|W|jm;+\rangle=\sum_{m''\mu}C^{jm'}_{1\mu\;jm''}
\langle 1^{-}\mu|a_{jm''}Wa^{\dagger}_{jm}|0^{+}\rangle.              \label{11}
\end{equation}
For the rotational scalar (\ref{10}), we have $m'=m$, and the collective contribution is
\begin{equation}
\langle jm;-|W|jm;+\rangle\approx \xi\sum_{\kappa m' \mu}(-)^{\kappa}C^{jm}_{1\mu\;jm'}\langle 1^{-}\mu|
D_{\kappa}|0^{+}\rangle\langle a_{jm'}|\sigma_{-\kappa}|a^{\dagger}_{jm}\rangle.  \label{12}
\end{equation}
In the form independent of projections, the necessary matrix element is given by
\begin{equation}
W_{10}=-\,\frac{\xi}{\sqrt{3(2j+1)}}\,(1^{-}||D||0^{+})(j\ell||\sigma||j\ell).   \label{13}
\end{equation}
Here the constant $\xi\equiv\xi_{j\ell}$ depends on the shell-model orbital of the odd nucleon.

\section{Estimating the interaction strength}

For the estimate by the order of magnitude of the effective ${\cal P}$- and ${\cal T}$-interaction
acting according to the suggested mechanism, we use the weak nucleon-nucleon interaction in the form
\cite{FKS}, where we neglect the velocity-dependent terms. The main contribution comes from
the contact interaction between the nucleons $a$ and $b$,
\begin{equation}
\hat{W}^{\circ}_{ab}=\,\frac{G}{\sqrt{2}}\,\frac{1}{2m}\,(\eta_{ab}\vec{\sigma}_{a}-\eta_{ba}
\vec{\sigma}_{b})\cdot\nabla_{a}\delta({\bf r}_{a}-{\bf r}_{b}).     \label{14}
\end{equation}
Here $G$ is the weak Fermi constant, $m$ the nucleon mass ($mc/\hbar$ in full units),
$\vec{\sigma}_{a,b}$ are spin operators and ${\bf r}_{a,b}$ are coordinate operators of interacting nucleons.
The strength of the effective interaction $W^{\circ}$ is regulated by the dimensionless parameters
$\eta_{ab}$ to be determined by future experimental measurements.

According to our idea, we limit ourselves to the search for the new effect related to the collective
dipole excitations. The earlier known collective mechanisms coming from the static octupole deformation
\cite{AFS96,SAF97}, soft octupole vibration \cite{FZ03}, or combined soft quadrupole and octupole modes
\cite{ZVA08} are the subject of further studies, theoretical and experimental. Here we are looking
for the collective dipole excitation of the even-even core. In eq. (\ref{14}) we consider the
interaction $W^{\circ}_{ab}$ of the valence nucleon $b$ with the nucleons $a$ in the core. Then the spin
operator in eq. (\ref{14}) should be related to the external nucleon in the shell-model
state $\psi_{b}$. The corresponding part of the effective interaction is given by
\begin{equation}
\hat{W}_{a}=-\,\frac{G}{\sqrt{2}}\,\frac{1}{2m}\,\eta_{ba}\left(\nabla_{a}\cdot\psi_{b}^{\ast}({\bf r}_{a})\vec{\sigma}_{b}\psi_{b}({\bf r}_{a})\right).                          \label{15}
\end{equation}
We are interested in the dipole component of this operator that takes part in the collective dipole
excitation of the core. To extract this component we can use the projection method \cite{FZ03}.

The contribution of the core nucleon $a$ to the collective dipole mode can be presented as a vector
\begin{equation}
\hat{{\bf W}}_{a}=C_{a}\hat{{\bf D}},                                 \label{16}
\end{equation}
where $\hat{{\bf D}}$ is the corresponding collective operator, such as the dipole moment or the Schiff
moment, while the effective amplitude $C_{a}$ for a given nucleon $a$ is determined by the projection,
\begin{equation}
C_{a}=\,\frac{(\hat{{\bf D}}|\hat{{\bf W}}_{a})}{(\hat{{\bf D}}|\hat{{\bf D}})},    \label{17}
\end{equation}
onto the normalized dipole state. As the amplitude of collective vibrations is small compared to the mean
nuclear radius, the normalization can be performed by integration over the unperturbed volume.

\subsection{Isovector dipole moment}

The simplest case corresponds to the isovector dipole excitation,
\begin{equation}
{\bf D}=\sum_{a}e_{a}{\bf r}_{a},                                            \label{18}
\end{equation}
with effective charges $e_{p}$ and $e_{n}$. The presence of the neutron skin in exotic nuclei can
enhance the low-energy dipole mode. Therefore we distinguish the proton and neutron equilibrium radii,
$R_{p}$ and $R_{n}$ (these are radii of equivalent uniform spheres). Then
\begin{equation}
(\hat{{\bf D}}|\hat{{\bf D}})=\,\frac{4\pi}{5}\,\left(e_{p}^{2}R_{p}^{5}+e_{n}^{2}R_{n}^{5}\right).
                                                                           \label{19}
\end{equation}

The overlap of the dipole operator (\ref{18}) with the weak interaction mechanism is given by
\begin{equation}
(\hat{W}_{a}|D_{z})=-\,\frac{G}{\sqrt{2}}\,\frac{1}{2m}\,e_{b}\eta_{ba}\int d^{3}r\,\psi^{\ast}_{b}({\bf r})
\vec{\sigma}\psi_{b}({\bf r})\cdot(\nabla z).                              \label{20}
\end{equation}
For a given orbital $(j\ell m)$ of the external nucleon $b$, we come to
\begin{equation}
(\hat{W}_{a}|D_{z})=-\,\frac{G}{\sqrt{2}}\,\frac{1}{2m}\,e_{b}\eta_{ba}\,\frac{1}{\sqrt{2j+1}}\,
C^{jm}_{jm\;10}(j\ell||\sigma||j\ell).                                   \label{21}
\end{equation}
Now we determine the projection amplitude (\ref{17}),
\begin{equation}
C_{a}[D]==-\,\frac{G}{\sqrt{2}}\,\frac{1}{2m}\,e_{b}\eta_{ba}\,\frac{3}{\sqrt{2j+1}}\,
C^{jm}_{jm\;10}(j\ell||\sigma||j\ell)\,\frac{1}{(\hat{{\bf D}}|\hat{{\bf D}})}.   \label{22}
\end{equation}
Comparison of this result with the general expression (\ref{13}) determines the effective constant
\begin{equation}
\xi[D]=\,\frac{G}{\sqrt{2}}\,\frac{1}{2m}\,\frac{15}{4\pi}\,\frac{e_{b}\eta_{b}}{e_{p}^{2}R_{p}^{5}
+e_{n}^{2}R_{n}^{5}},                                                       \label{23}
\end{equation}
where $\eta_{b}$ is the constant $\eta_{ba}$ averaged over the core nucleons with the weights $N/A$
and $Z/A$ for neutrons and protons, respectively. In this approximation the only dependence on
the unpaired orbital in the odd nucleus can enter through $\eta_{b}$.

Collecting everything (consider $\alpha$ real) we obtain
\begin{equation}
\langle S\rangle_{0}=-2\alpha\langle j,m=j;+|S_{\kappa=0}|j,m=j;-\rangle=\,\frac{2\alpha}{\sqrt{3}}
\,C^{jj}_{10\;jj}\,(0^{+}||S||1^{-}),                                          \label{24}
\end{equation}
where
\begin{equation}
\alpha=\,\frac{W_{10}}{\Delta E}=-\,\frac{1}{\Delta E}\,
\frac{\xi}{\sqrt{3(2j+1)}}\,(1^{-}||D||0^{+})(j\ell||\sigma||j\ell).     \label{25}
\end{equation}
Taking $\xi=\xi[D]$ from eq. (\ref{23}) and generalizing for the case of several excited states $|1^{-}_{i}\rangle$
with excitation energies $\Delta E_{i}$,
\begin{equation}
\alpha=-\,\frac{1}{\sqrt{3(2j+1)}}\,\frac{G}{\sqrt{2}}\,\frac{1}{2m}\,\frac{15}{4\pi}\,
\frac{e_{b}\eta_{b}(j\ell||\sigma||j\ell)}{e_{p}^{2}R_{p}^{5}+e_{n}^{2}R_{n}^{5}}\,\sum_{i}
\frac{(1^{-}_{i}||D||0^{+})}{\Delta E_{i}}.     \label{26}
\end{equation}
The last matrix element in (\ref{26}) is
\begin{equation}
(j\ell||\sigma||j\ell)=\left[j(j+1)-\ell(\ell+1)+\,\frac{3}{4}\right]\sqrt{\,\frac{2j+1}{j(j+1)}\,}.
                                                                       \label{27}
\end{equation}
The Clebsch-Gordan coefficient in eq. (\ref{24}) equals to $\sqrt{j/(j+1)}$.

\subsection{Schiff moment}

The same techniques can be used for the choice of the Schiff moment (\ref{6}) as the dipole operator
(\ref{16}) acting with the weak interaction. The normalization (\ref{19}) can be found now as
\begin{equation}
(\hat{{\bf S}}|\hat{{\bf S}})=\,\frac{8}{315}\,4\pi e^{2}R_{p}^{9}x^{2}.  \label{28}
\end{equation}
The projection of the weak interaction, similarly to eq. (\ref{20}), is expressed as
\begin{equation}
(\hat{W}_{a}|S_{z})=-\,\frac{G}{\sqrt{2}}\,\frac{x}{2m}\,e_{b}\eta_{ba}\int d^{3}r\psi^{\ast}_{b}({\bf r})
\vec{\sigma}\psi_{b}({\bf r})\cdot\nabla\Bigl((r^{2}-(5/3)\langle r^{2}\rangle_{{\rm ch}})z\Bigr).  \label{29}
\end{equation}
Instead of (\ref{21}) we now have
\begin{equation}
(\hat{W}_{a}|S_{z})=-\,\frac{G}{\sqrt{2}}\,\frac{x}{2m}\,e_{b}\eta_{ba}\,\frac{1}{\sqrt{2j+1}}\,
C^{jm}_{jm\;10}(j\ell||F||j\ell),                                   \label{30}
\end{equation}
where the ${\cal P}$- and ${\cal T}$-odd vector operator is introduced
\begin{equation}
{\bf F}=2(\vec{\sigma}\cdot{\bf r}){\bf r}+\vec{\sigma}\left(r^{2}-\frac{5}{3}\,
\langle r^{2}\rangle_{{\rm ch}}\right).                         \label{31}
\end{equation}
The projection amplitude is now slightly more complicated,
\begin{equation}
C_{a}[S]==-\,\frac{G}{\sqrt{2}}\,\frac{1}{2m}\,e_{b}\eta_{ba}\,\frac{3}{\sqrt{2j+1}}\,
C^{jm}_{jm\;10}(j\ell||F||j\ell)\,\frac{1}{(\hat{{\bf S}}|\hat{{\bf S}})},   \label{32}
\end{equation}
which leads to the effective interaction constant
\begin{equation}
\xi[S]=\,\frac{G}{\sqrt{2}}\,\frac{1}{2m}\,\frac{945}{8x}\,\frac{1}{4\pi}\,\frac{e_{b}\eta_{b}}
{e_{p}^{2}R_{p}^{9}}\,\frac{(j\ell||F||j\ell)}{(j\ell||\sigma||j\ell)}.      \label{33}
\end{equation}

The matrix element of the operator ${\bf F}$ can be easily found with standard methods,

\begin{equation}
\frac{(j\ell||F||j\ell)}{(j\ell||\sigma||j\ell)}=\,\frac{j+2}{j+1}\langle r^{2}\rangle_{j\ell}-
\,\frac{5}{3}\,\langle r^{2}\rangle_{{\rm ch}}, \quad j=\ell+\,\frac{1}{2},            \label{34}
\end{equation}
and
\begin{equation}
\frac{(j\ell||F||j\ell)}{(j\ell||\sigma||j\ell)}=\,\frac{j-1}{j}\langle r^{2}\rangle_{j\ell}-
\,\frac{5}{3}\,\langle r^{2}\rangle_{{\rm ch}}, \quad j=\ell-\,\frac{1}{2}.            \label{35}
\end{equation}
In the case of the $p_{1/2}$ level, eq. (\ref{35}), the effect is enhanced.
 
\section{Resulting Schiff moment}

The final result for the expectation value of the Schiff moment in the ground state $|jm\rangle_{0}$
of the odd nucleus,
\begin{equation}
\langle S\rangle_{0}=_{0}\langle j,m=j|S_0|j,m=j\rangle_{0},                 \label{36}
\end{equation}
can be expressed as
\begin{equation}
\langle S\rangle_{0}=\,\frac{2}{3}\,\sqrt{\,\frac{j}{(j+1)(2j+1)}\,}\,\xi[S](j\ell||\sigma||j\ell)
\sum_{i}\frac{|(1^{-}_{i}||S||0^{+})|^{2}}{\Delta E_{i}}.                \label{37}
\end{equation}
Here we have written the answer for the choice of ${\bf S}$, Sec. 4.2, as an operator ${\bf D}$ exciting the collective mode.

In spite of a relatively large energy denominators, we can expect a
result that is comparable with what has been found for the pure single-particle excitations which, in fact, usually have the excitation energy of the same order of magnitude. This is clear from the resulting
expression (\ref{37}) if there is a coherent low-lying pygmy mode that supports an appreciable
fraction of the dipole sum rule. The collective character of the pygmy-mode is expected to
be more pronounced if its centroid is already in the continuum because of the additional coherent interaction
of particle-hole states through common decay channel(s) \cite{Szagreb}. If there is no RPA-type collectivization
through continuum coupling but the amplitudes $\xi[S]$ are of the same sign for several partial
excitations, they will coherently add to the amplification of the Schiff moment.

\section{Quantitative estimate}

We now present an example of a numerical evaluation of the Schiff moment resulting from the isoscalar dipole strength. Our purpose is mainly to obtain an order of magnitude estimate for $S$.

We estimate  the expression (\ref{37}) for a nucleus with $A=201$, with the extra nucleon occupying the $p_{1/2}$ orbital. For the strength distribution of the ISDR we use the results of the calculation contained in ref.
\cite{kvasil03}. This calculation shows that the strength is concentrated in two broad peaks, one at lower energy (the  centroid around 10 MeV in $A\sim 200$ nuclei) and the higher energy peak (centroid around 20 MeV).  Characterizing the results one can say that about $10^5$  of the strength (in units of fm$^6$) is contained in the lower peak and about $2\cdot 10^5$ fm$^6$ in the upper peak. For our estimate of summation in eq. (\ref{37}) we introduce the total strength and corresponding energy centroids for the two peaks; for the radius appearing in eq. (\ref{33}) we use $R=6$ fm.
Collecting all numbers we find for the Schiff moment the result
$S=0.8\cdot 10^-8 \eta\,  e$ fm$^3$. This is of the same size as the Schiff moments evaluated in single-particle models \cite{dmitriev04,dobaczewski05}, including effects of the core polarization \cite{dmitriev05}.

It would be of interest to perform a more detailed microscopic calculation by calculating explicitly the inverse energy weighted sum in the framework of the random phase approximation. The inverse energy weighted sum is equivalent to the polarization of the nucleus that can be computed in the constrained Hartree-Fock scheme with the Schiff operator being the constraint. An even more significant possibility is to calculate the ISDR in deformed nuclei or nuclei away from the stability line where one could expect the lowering of strength leading to an enhancement of
the Schiff moment. A similar estimate for the isovector dipole resonance, eq. (\ref{26}), is not possible since this expression requires the amplitudes of the isovector dipole rather then the strength. However in a fully microscopic approach one would be able to calculate the Schiff moment for this mechanism as well. In this case some of the contributions of the low-lying pygmy dipole resonance might become significant.

\begin{center}
{\bf ACKNOWLEDGEMENTS}
\end{center}

The work was supported by the NSF grant PHY-1068217. The authors thank National Superconducting Cyclotron Laboratory
at Michigan State University (N.A.) and Tel Aviv University (V.Z.) for hospitality. The collaboration with R.A. Sen'kov at the beginning of this work is gratefully acknowledged. N.A. is thankful to Ch. Stoyanov for helpful discussions.

\end{document}